Title: The Dynamic Role of Aerosol and Exudate Transport in the Diffusion of Lung Infection in Respiratory Infectious Diseases (taking SARS-CoV-2 as an example): A Hypothesis Model

Shi Qiru*

* Correspondence: shiqiru@lzcu.edu.cn

Lanzhou City University

11 Jiefang Road, Anning District, Lanzhou, Gansu, China

## 1. Abstract

This paper proposes a hypothetical model for the dual role of respiratory aerosols and inflammatory exudates in the dynamics and progression of SARS-CoV-2 lung infection. Starting from a new paradigm in infectious disease transmission, we reflect on the often-overlooked role of physical transmission media within the host individual. The hypothesis posits that tiny aerosols (including those inhaled externally and those self-generated and re-inhaled by the host) play a crucial role in the initial seeding and early expansion of the infection in the lungs, explaining the multifocal characteristics observed in early CT imaging. As the infection progresses, inflammatory exudates, formed due to lung inflammation, become a new efficient vehicle, driving the large-scale spread of the virus within the lungs and accounting for the development of diffuse lesions. This model reveals a "dynamic equilibrium point" where the dominant mechanism shifts from aerosol-mediated to exudate-mediated spread. Although direct validation of this hypothesis faces ethical and technical challenges, existing clinical imaging, viral kinetics, and epidemiological patterns provide indirect support. The paper also conceptualizes ideal experimental designs and retrospective analyses to validate the hypothesis. Finally, we discuss the implications of this hypothesis for public health practice, emphasizing the importance of improving ventilation in the microenvironment of infected individuals to achieve a "for all, by all" (literally "everyone for me, I for everyone") bidirectional protection. This research aims to provide a new framework for understanding the pathophysiology of respiratory infectious diseases and to offer theoretical basis for developing more cost-effective and broadly applicable intervention strategies.

## 2. Introduction: The Enigma of Lung Infection in the New Paradigm of Respiratory Infectious Disease Transmission

In recent years, with the deepening understanding of respiratory infectious disease transmission mechanisms, particularly the updated and widely accepted definition of airborne transmission (i.e., aerosol transmission) by authoritative organizations such as the World Health Organization (WHO), a significant scientific paradigm shift is underway. This shift leads us to no longer solely focus on large droplets and direct contact, but to more broadly recognize the critical role of tiny suspended particles in the air (aerosols) in disease transmission between individuals. However, this cognitive breakthrough immediately raises a deeper, yet often overlooked question: When pathogen-containing aerosols transmit between individuals, do airborne aerosols, including virus-laden aerosols exhaled and then re-inhaled by the host, also influence or even dictate viral amplification and disease progression within the host individual, particularly in the lungs (despite the fact that, from the perspective of pathogen quantity, the number of pathogens deposited in the lungs via aerosol inhalation is negligible compared to their exponential amplification within the host)? This might not be a far-fetched idea.

The pathways by which viruses enter the host lung and establish infection foci can be diverse, including but not limited to: inhalation of virus-laden aerosol particles, systemic dissemination from other body parts via blood or lymphatic systems (hematogenous or lymphatic dissemination), and direct spread of the virus between infected cells. However, considering the ability of tiny aerosol particles to penetrate deep into the alveoli and deposit randomly, along with the multifocal characteristics observed in early CT imaging of lung infections, this paper will focus on the dynamic role of pathogen-containing aerosols (both externally inhaled and self-generated and re-inhaled by the host) in establishing initial infection foci and driving the early expansion of infection in the lungs. This focus is chosen partly because aerosols, as a physical medium, have dynamic behaviors in the lung microenvironment that can be analyzed, modeled, and validated; more importantly, compared to routes like systemic dissemination that are difficult to directly intervene with environmental measures, improving ventilation and air filtration to reduce aerosol exposure and re-inhalation holds practical public health intervention potential, thus giving research into aerosol mechanisms higher translational application value and relatively higher research feasibility.

For a long time, infectious disease research has often fallen into a certain thought inertia in understanding pathogen transmission routes and in-vivo dissemination mechanisms. Looking back at history, during the cholera epidemic in the 19th century, the mainstream scientific community was fixated on the "miasma theory," neglecting the more fundamental transmission medium of contaminated water sources. It was not until John Snow's groundbreaking work that the truth of water-borne fecal-oral transmission was gradually revealed, laying the foundation for effective cholera prevention and control. Similarly, for respiratory infectious diseases, current research may overemphasize the efficiency of viral intracellular replication after entering the host and its transmission chain among populations, while relatively neglecting how viruses in the external environment (especially aerosols) continuously influence disease progression within the host individual, particularly in the complex and vast pulmonary microenvironment, and how the host's own physiological changes (such as inflammation and fluid exudation) alter the medium for viral diffusion within the lungs.

Specifically for COVID-19 pneumonia caused by SARS-CoV-2, although CT imaging clearly shows the dynamic evolution of lung lesions from focal to diffuse [1], a unifying theory is lacking to precisely explain this rapid and often randomly distributed spread pattern. Traditional views tend to attribute the expansion of lung lesions mainly to sequential local spread of the virus between infected cells, or emphasize the role of host immune response dysregulation (e.g., cytokine storm) leading to widespread tissue damage and escalating inflammation in driving the disease course. These perspectives undoubtedly shed light on critical aspects of the pathological process; for example, immune dysregulation indeed leads to increased pulmonary microvascular permeability and the formation of large amounts of inflammatory exudate, which itself signifies lung injury and severely impairs gas exchange. However, traditional explanations may not have sufficiently focused on how these pathophysiological changes (especially the formation of exudate) transform into a "medium and driving force" for large-scale, rapid physical dissemination of the virus within the lung. Specifically, they might underestimate the potential for continuous inhalation of infectious sources or self-exhaled high-concentration virus aerosols to seed new infection foci in the lungs, and the critical role of these exudates themselves, when inflammation leads to their accumulation in airspaces, as an efficient physical transport vehicle promoting the rapid, long-distance spread of viral suspensions between lung segments and even lobes, driven by gravity, respiratory airflow disturbances, etc.

This paper aims to propose a hypothetical model for the dynamic diffusion of respiratory infectious diseases (taking SARS-CoV-2 as an example) within the host lung. We will delve into the unique anatomical and physiological characteristics of the lung, and on this basis, construct a dynamic equilibrium model that distinguishes between initial seeding mediated by external aerosols, early expansion mediated by self-generated aerosols, and large-scale dissemination dominated by exudate transport after inflammation. It should be noted that, to highlight the role of physical transmission media, this model currently simplifies the complex influence of the host immune system on pathogen amplification and clearance. Through this novel perspective, we hope to provide a new framework for understanding the pathophysiology of respiratory infectious diseases and to offer a theoretical basis for formulating lower-cost, broad-spectrum, and more practical prevention and control strategies, such as emphasizing the importance of improving ventilation in the microenvironment of infected individuals.

3. Aerosols and Respiratory Lung Anatomical-Physiological Characteristics

The lung, as the core organ of the respiratory system, has unique anatomical structures and physiological functions that determine its exposure patterns to airborne particles (including viral aerosols) and its response to infection. Understanding these characteristics is fundamental to constructing a hypothesis of lung viral diffusion.

3.1 Deep Pulmonary Deposition of Tiny Aerosols

The respiratory system is a highly branched network of tubes, extending from the nasal cavity, pharynx, through the trachea and various levels of bronchi, finally reaching the alveoli of the lungs. Aerodynamic studies show that aerosol particles of different sizes deposit in different parts of the respiratory tract. Larger droplets (diameter > 5-10 μm) usually deposit in the upper respiratory tract (e.g., nasal cavity, pharynx) due to inertial impaction. In contrast, tiny aerosol particles (diameter < 5 μm), especially sub-micron and even nano-sized particles, due to their lower sedimentation velocity and stronger Brownian motion, can effectively bypass the clearance mechanisms of the upper respiratory tract, penetrate deep into the lower respiratory tract, and even directly deposit in the small bronchioles and alveolar regions of the lungs (although the deposition percentage in these regions is significantly lower than in the nose, pharynx, and trachea [2]). This deep deposition characteristic makes tiny aerosols an important vehicle for viruses to enter the lung parenchyma and establish initial infection foci. For SARS-CoV-2, its diameter is approximately 80-120 nanometers (nm), falling within the typical range of tiny aerosols capable of deep deposition.

3.2 Unique Spatial, Cellular, and Exudate Film Characteristics of the Lung

The unique internal environment of the lung is crucial for early viral attachment and spread:

3.2.1 Vast Surface Area

The total surface area of an adult lung, when unfolded, can reach approximately 50-100 square meters, equivalent to the size of a tennis court. This enormous surface area is designed to maximize gas exchange efficiency and also provides a vast potential deposition and infection interface for airborne pathogens. From the perspective of pathogen transmission, the large alveolar surface significantly increases the probability of pathogen contact with host cells; however, this characteristic also means that a small number of initially inhaled pathogens, lacking effective diffusion mechanisms, tend to be highly concentrated (in the absence of exogenous transport vehicles, pathogen diffusion is limited by the low fluidity of the alveolar microenvironment) but clustered in extremely small areas and distributed sparsely. Furthermore, compared to the higher viral loads in the oral and nasal pharynx, the initial pathogen dose inhaled into the lungs is usually lower, and must overcome multiple innate immune defense mechanisms, such as the mucociliary clearance system and alveolar macrophage phagocytosis, to establish an effective infection in the lungs. Therefore, the process by which pathogens colonize the lung and initiate sustained amplification requires overcoming the dual biological limitations of sparse spatial distribution and host immune clearance.

3.2.2 Hierarchical Branching and Tortuous Structure of the Airway Tree: A Natural Physical Barrier

In addition to the vast terminal surface area, the lung's conducting airway system (from the trachea and main bronchi to various levels of bronchioles) exhibits a complex, multi-level bifurcating and tortuous tree-like structure. This unique anatomical configuration acts as an important physical barrier aerodynamically.

Interception of large particles: When inhaled airflow carries pathogen-laden particles through these bends and bifurcations, due to inertia, larger particles (typically referring to aerodynamic diameters greater than 5-10 micrometers, such as droplets or large dust particles) are more likely to impact and deposit on the mucosal lining of the upper airway walls when the airflow direction changes. These captured particles can then be transported upwards by the mucociliary clearance system to the pharynx for coughing out or swallowing, thus effectively preventing them from penetrating deeper into the more sensitive bronchioles and alveolar regions of the lung.

Partial restriction of early exudate spread during infection: If the upper respiratory tract or larger bronchi are infected first and produce inflammatory exudate, the branching and gravitational forces of the airways can, to some extent, help confine early-formed, small amounts of exudate or larger droplets to the local or upper airway sections where the infection occurred, slowing their diffusion deeper into the lung by gravity alone or minor airflow disturbances.

However, it is noteworthy that for tiny aerosol particles (typically

4. SARS-CoV-2 Dynamic Infection Hypothesis Model in the Lung

The clinical manifestations of COVID-19 pneumonia, especially the rapid evolution of lung lesions from focal to diffuse in CT imaging, suggest a complex diffusion mechanism of the virus within the host lung that goes beyond mere cell-to-cell replication and immune response. This section will propose a dynamic hypothetical model explaining how SARS-CoV-2 utilizes aerosols as a vehicle for initial seeding and early expansion, and how inflammation-induced exudates transform into the dominant medium for exudate transport, thereby achieving large-scale dissemination of the virus within the lung.

4.1 Phase One: Aerosol-Dominant Initial Seeding and Early Expansion

In the early stage of infection, the establishment of lung infection foci is a critical starting point for disease progression. While viruses can reach the lungs and establish initial infection foci through various routes (e.g., hematogenous dissemination as mentioned in the introduction), this hypothesis suggests that in the early stages of disease, inhaled aerosol particles (whether from external sources or self-generated by the host) are the primary physical seeding carriers leading to multiple discrete infection foci in the lungs, due to their ability to directly deliver viruses deep into the lung and form randomly distributed deposition sites. At this time, the lung environment is relatively "dry," inflammation has not yet significantly started, and the normal mucociliary clearance system is relatively intact. Early viral spread in the lung primarily relies on physical aerosol deposition and subsequent local replication.

4.1.1 External Aerosol-Mediated Initial Seeding (Continuous Process)

When the host inhales external tiny aerosols containing SARS-CoV-2 from an infected source, these particles can penetrate the upper respiratory tract, enter the small bronchioles and alveolar regions of the lungs, and deposit [2]. As mentioned earlier, successful deep lung infection depends on inhaling sufficiently high concentrations of viral aerosols and adequate exposure time. It is important to note that this external aerosol exposure is not a one-time event but can be a continuous process, especially in certain high-risk exposure scenarios:

Individuals sharing confined spaces for extended periods at close range: For example, roommates in dormitories who study and live together during winter, or family members living together, are continuously exposed to each other's exhaled viral aerosols due to prolonged presence in poorly ventilated, confined spaces.

High-concentration viral aerosol exposure environments: Certain activities significantly increase the viral load in exhaled aerosols and the risk of transmission. For instance, studies have shown that the concentration of SARS-CoV-2 aerosols exhaled by infected individuals while singing can be significantly higher than during quiet breathing, with infectivity sufficient to transmit COVID-19 within minutes. Similar high-risk scenarios include strenuous exercise, loud speaking, or coughing, all of which produce high-concentration, highly infectious aerosols, leading to a rapid increase in environmental viral load [13].

In these situations, the seeding of external aerosols in the host lung can be a continuous, low-level process, constantly introducing new infection foci or enhancing existing ones in the lungs. This also explains why infections often cluster in certain co-exposed populations, and individual disease progression may be related to the duration and intensity of exposure.

4.1.2 Self-Aerosol-Mediated Early Intrapulmonary Expansion (Driven by Oropharyngeal Viral Load Peak)

External aerosol-mediated initial seeding is the starting point of lung infection and can continue under sustained exposure. However, as discussed in the previous section on lung anatomical and physiological characteristics, early viral amplification in the lung is relatively limited. The thin fluid layer on the alveolar surface and the mucociliary clearance system do not favor rapid viral spread on the epithelial surface.

Epidemiological and viral kinetics studies show that SARS-CoV-2 viral load in the infected individual's oropharynx typically peaks within the first 1 to 5 days after symptom onset. This high viral load period significantly increases the infected individual's ability to exhale high-concentration viral aerosols, thereby rapidly increasing the viral load in the surrounding air. Simultaneously, the host's immune response (especially adaptive immune response) exhibits significant delay. Typically, from virus entry into the host to the production of effective specific antibodies or cellular immunity, several days or even longer are required (this delay will be even more pronounced when the lung viral load is extremely low and confined to a very small area). It is within this window of relatively weak immune response, which has not yet rapidly triggered large-scale inflammation and fluid exudation, that high-load viral aerosols exhaled by the self are re-inhaled, becoming a key driving force for the expansion of infection within the lungs.

This time lag between the peak oropharyngeal viral load and the peak pulmonary viral load reveals a crucial transmission mechanism: rapid viral replication in the oropharyngeal region and the resulting high concentration of self-generated aerosols can lead to new viral particles depositing into other lung regions with each respiratory cycle. This continuous, high-frequency

"self-reseeding" effectively overcomes the problem of inefficient early local viral spread in the lungs, by repeatedly establishing new infection foci in different lung regions. This aligns highly with the observation that the peak viral load in the lungs typically occurs between 5 to 10 days after symptom onset [1, 8], suggesting that early oropharyngeal high viral loads leading to self-aerosol reseeding are very likely the main cause of subsequent widespread viral amplification and peak viral load in the lungs. It allows the virus to rapidly expand the infection area in the lungs in the early stages, before the immune system has fully initiated inflammatory responses and fluid exudation, thereby limiting exudate-mediated diffusion, laying the foundation for subsequent disease progression.

4.2 Phase Two: Inflammation and Exudate Formation, Exudate Transport Dominates Large-Scale Dissemination

The efficiency of viral dissemination within the host body is largely influenced by the physicochemical properties of the fluid vehicle it relies on. For example, in the early stages of infection, viral kinetics in different parts of the upper respiratory tract may vary due to differences in the local microenvironment: saliva in the oral cavity has high fluidity, which may provide more favorable conditions for local viral dissemination and efficient amplification on mucosal surfaces, thereby promoting early high viral load formation; in contrast, secretions in the nasal cavity may be relatively viscous or have weaker fluidity, which may affect the local viral diffusion and amplification rate [9, 10, 11, 12]. This example suggests that the characteristics of the fluid vehicle are a critical factor in understanding viral transmission dynamics.

When the infection progresses to the second phase, with significant inflammatory response in the lungs accompanied by the formation of large amounts of exudate, this virus-rich inflammatory exudate constitutes a new, more efficient viral vehicle. At this point, the exudate accumulated within the alveolar and bronchiole lumens, by virtue of its fluidity and virus-carrying capacity, replaces the early aerosol-dominated dissemination mode, becoming the core medium and dominant mechanism driving large-scale, rapid diffusion of the virus within the lung parenchyma and leading to the formation of diffuse lesions.

4.2.1 Inflammation-Induced Fluid Exudation and Microenvironmental Changes

Extensive viral replication within the alveoli and airways, along with a severe host immune inflammatory response (e.g., cytokine storm), leads to increased permeability of pulmonary capillaries. Large amounts of plasma, inflammatory cells, and proteins exude, forming widespread exudates and inflammatory fluid accumulation within the alveoli and airways. Additionally, damage to alveolar epithelial cells and impairment of ciliary clearance further prevent these fluids and captured viruses from being effectively cleared. These accumulated fluids not only alter the physical microenvironment of the lung but also provide a new, highly efficient medium for viral diffusion.

4.2.2 Exudate-Mediated Viral Transport and Large-Scale Spread

Once large amounts of inflammatory exudate and fluid accumulate in the lungs, these fluids serve as an ideal vehicle for extensive surface diffusion of the virus within the lung. Viruses can dissolve or suspend in these fluids and achieve large-scale spread through the following mechanisms:

Exudate production and local expansion: As inflammatory exudate accumulates within the alveoli and airways, these fluids provide a medium for viral diffusion, aiding the spread of the virus within fluid-filled regions to surrounding areas, thereby expanding the lesion area.

Airway branching connectivity: The highly branched structure of the lung airways, when filled with fluid, acts like a network of "rivers" providing physical channels for the virus to diffuse in the fluid from one lesion to adjacent or more distant airways and alveolar units.

Diffuse lesion formation: This exudate-mediated diffusion mechanism can lead to the rapid spread and fusion of viruses from initial discrete infection foci to the surrounding environment, ultimately forming the bilateral, diffuse ground-glass opacities and consolidation observed on CT images.

4.3 Dynamic Equilibrium Point: Shift in Dominant Transmission Mechanism

This hypothesis suggests that lung viral diffusion has a dynamic equilibrium point ($T_{threshold}$), which marks the shift in the dominant diffusion mechanism:

Early stage ($t < T_{threshold}$): Lung inflammation is mild, and fluid exudation is not significant. At this time, $P_{aerosol}(t)$ is high, including initial seeding by external aerosols and early intrapulmonary reseeding by self-aerosols; $P_{exudate}(t)$ is low. Viruses primarily establish discrete infection foci in the lungs via aerosols.

Later stage ($t \geq T_{threshold}$): Lung inflammation intensifies, and large amounts of exudate form. At this time, the influence of $P_{aerosol}(t)$ on further intrapulmonary diffusion decreases, while $P_{exudate}(t)$ increases. Viruses primarily achieve large-scale spread and lesion fusion in the lungs through exudate-mediated surface diffusion.

$T_{threshold}$ can be a time threshold that varies depending on the infection situation, or it can be a threshold of inflammation severity or infection extent corresponding to the shift from focal to diffuse lesions in CT imaging. This shift reflects how host physiological responses (inflammation) in turn shape the physical mode of pathogen diffusion within the body.

5. Hypothesis Validation: Indirect Evidence and Challenges of Direct Verification

Directly validating the "dynamic role of aerosol and exudate transport in the diffusion of respiratory infectious disease lung infection" hypothesis faces significant challenges, primarily due to limitations in experimental conditions, technological means, and ethical considerations. However, despite the difficulties of direct validation, we can still support this hypothetical model through a series of indirect evidence.

5.1 Challenges of Direct Verification

Directly verifying the specific mechanisms of this hypothesis in living human subjects (e.g., real-time tracking of specific aerosol particle deposition in the lungs, or quantifying virus diffusion rates mediated by fluid flow) faces immense ethical and technical challenges:

Ethical limitations: Conducting experiments in humans that involve deliberate exposure to viruses or invasive real-time monitoring presents severe ethical barriers and cannot be performed.

Difficulty of in-vivo real-time tracking: Real-time, dynamic observation of virus particles (nanometer scale) or microscopic viral movement in fluids (micrometer-millimeter range) within the complex three-dimensional structure of living human lungs is almost impossible. Existing imaging technologies (e.g., CT, MRI) lack the resolution to capture these microscopic dynamic processes.

Challenges in simulating complex microenvironments: The lung is a highly dynamic and complex microenvironment, involving multiple factors such as respiratory movements, airflow, fluid flow, mucociliary movement, and immune cell activity. In vitro models (e.g., organoids, microfluidic chips) or animal models cannot fully and accurately replicate these complex dynamic processes in human lungs, especially aerosol deposition under different breathing patterns and the flow of inflammatory exudates in the alveolar-airway network.

Distinguishing different diffusion mechanisms: During infection progression, multiple mechanisms such as direct cell-to-cell spread, aerosol re-deposition, fluid-mediated diffusion, and immune cell-mediated transport may occur simultaneously. How to isolate and quantify the contribution of each mechanism in in-vivo experiments is a tremendous technical challenge.

Dynamics of viral load and distribution: Viral load and distribution are highly dynamic in the lungs, making accurate sampling and analysis at different time points extremely difficult to capture evidence of mechanism shifts.

5.2 Supporting Indirect Evidence

5.2.1 Consistency with Clinical, Viral Kinetic, and Epidemiological Observations

This hypothetical model is consistent with many observed phenomena of COVID-19, which provide indirect support:

Clinical imaging and pathological observations: Early CT images of the disease show multifocal, discrete ground-glass opacities rather than a single center of spread, suggesting that the virus initiates infection through multiple "seeding" points, consistent with the random deposition characteristics of aerosols. Subsequent rapid lesion fusion and diffuse progression, as well as the non-uniform, patchy lesion distribution and diffuse alveolar damage pathologically [1], are consistent with the hypothetical phase where the virus undergoes large-scale surface diffusion mediated by inflammatory exudates.

Viral kinetics: The phenomenon of SARS-CoV-2 viral load peaking in the oropharynx earlier than in the lungs indirectly supports the inference that high-load aerosols generated in the oropharynx and re-inhaled by the self drive the early expansion of infection in the lungs, leading to a subsequent peak in viral load.

Epidemiological patterns: The association between prolonged close contact in poorly ventilated confined spaces (including activities by infected individuals that generate high-concentration aerosols, such as talking or singing) and high infection risk and disease severity also corroborates the hypothesis's critical role of aerosols (whether from external sources or self-generated) in early lung infection [13].

5.2.2 Conceptual Design of an Ideal Experiment to Validate the Hypothesis

Given the difficulty of real-time observation directly in humans, a finely designed comparative experiment might be needed to more directly validate the key role of "self-aerosol-mediated early intrapulmonary expansion" in this hypothesis. The following is a conceptual design for an ideal experiment aimed at evaluating the impact of self-aerosol re-inhalation on disease progression by controlling external aerosol exposure levels:

Cohort Selection and Grouping: Select a relatively stable population, such as one or several classes in a school, especially during peak viral prevalence (e.g., winter), and randomly divide them into an intervention group and a control group.

Control of External Aerosol Transmission (Intervention Group): Implement strict 24-hour air purification control in the indoor environment of the intervention group, including but not limited to:

Significantly increasing mechanical ventilation rates.

Using high-efficiency air filtration systems (e.g., HEPA filters) to ensure extremely low external viral aerosol loads in the air.

Control Group: The control group maintains standard environmental ventilation and human interaction patterns. Alternatively, if conditions permit and high-quality historical data are available, the intervention group could be compared with historical control data from populations with similar demographic characteristics and exposure backgrounds.

Behavioral Control: During the study period, all participants are asked to avoid activities known to produce extremely high concentrations of aerosols (e.g., prolonged singing). Simultaneously, participants are encouraged to maintain good oral hygiene, such as frequent gargling and mouth rinsing, to minimize the peak viral load of aerosols generated in the oral cavity and pharynx.

High-Frequency Monitoring: Conduct high-frequency pathogen infection monitoring (e.g., for SARS-CoV-2) for all participants in both groups, such as daily or every two days antigen tests, or more frequent nucleic acid tests, to minimize the omission of asymptomatic or mildly symptomatic infected individuals and ensure accurate assessment of infection rates and onset times.

Outcome Evaluation: Compare the cumulative infection rates (mechanical ventilation and air filtration systems may affect infection rates), disease severity after symptom onset (e.g., progression to pneumonia, hospitalization rates), and characteristics and progression speed of lung imaging changes between the two groups during the study period.

Expected Results and Hypothesis Association: If the disease progression and severity in the intervention group (where total viral aerosol exposure is significantly reduced through strict control of the external environment) are significantly lower than in the control group (where viral aerosol exposure is under conventional environmental conditions), then this would strongly support the hypothesis regarding the important role of environmental viral aerosols (including both external inhalation and self-generated aerosols accumulating and being re-inhaled) in driving and aggravating lung infection progression.

5.2.3 Retrospective Associative Analysis Based on Historical Cohort Data:

In addition to prospective ideal experimental designs, retrospectively analyzing historical cohort data that meet specific conditions can also provide indirect correlational evidence for this hypothesis. The logic of this analysis is similar to the "ideal experimental design concept" (4.2.2) mentioned above, namely, by comparing different human cohorts who, under similar backgrounds, might have significant differences in total viral aerosol exposure due to environmental factors (especially air quality control measures), and observing whether there are differences in their disease progression and severity after infection.

5.2.3.1 Ideal Historical Data Cohort Characteristics and Analysis Concept:

Cohort Selection and Grouping Analogy: Look for historically (e.g., during specific phases of the COVID-19 pandemic) two or more large human cohorts with similar characteristics (e.g., age structure, underlying health conditions, vaccination status, similar geographic locations and prevalent strains) but who were long-term exposed to different indoor air environments. For example:

Schools, office buildings, or specific communities that implemented strict and continuous air purification measures (e.g., high-efficiency mechanical ventilation, HEPA filtration) compared to control populations that did not implement similar measures.

Populations within the same institution or community where air quality differed significantly long-term due to architectural design or management strategies.

Indirect Assessment of Environmental Aerosol Exposure Levels: While direct measurement of aerosol viral load in historical data is unlikely, the relative exposure levels of viral aerosols in different cohorts' environments can be indirectly inferred through recorded environmental parameters (e.g., ventilation rates, air changes per hour, type and operating time of air purification equipment, per capita space area, etc.).

Consideration of Behavioral Control: If historical data can provide information on whether the cohort population generally adopted behaviors to avoid high aerosol production (e.g., large-group singing events) or enhanced oral hygiene measures during specific periods, this would help reduce confounding factors.

Alternatives and Calibration for High-Frequency Monitoring: Ideal historical data should include as comprehensive infection monitoring records as possible, for example:

Systematic, routine nucleic acid testing or antigen screening data, to more accurately capture true infection incidence and reduce omissions of mild or asymptomatic infections due to reliance on symptom reporting.

Serological survey data, to assess cumulative infection rates and prior infection status.

5.2.3.2 Outcome Evaluation: The core is to compare, in these cohorts whose total viral aerosol exposure might differ due to environmental differences:

Adjusted infection rates (correcting for differences in testing frequency and sensitivity as much as possible).

Disease severity after infection onset (e.g., incidence of pneumonia, hospitalization rates, severe disease rates, mortality rates).

If imaging data are available, compare the extent, progression speed, and characteristics of lung lesions.

5.2.3.3 Expected Results and Hypothesis Association:

If, in historical cohorts where total viral aerosol exposure was significantly reduced through strict control of the external environment (e.g., cohorts consistently in highly ventilated and filtered environments), a significant slowdown in disease progression and reduction in severity is observed (after adjusting for possible initial infection rate differences), then this would provide important indirect support for this hypothesis. Specifically, it would suggest that environmental viral aerosols (including the portion inhaled from the external environment and the portion self-generated, accumulated locally, and re-inhaled) play an important role in driving and exacerbating lung infection progression.

5.2.3.4 Challenges and Limitations:

The main challenges in conducting such retrospective analyses of historical data are:

Data Availability and Quality: It is extremely difficult to obtain high-quality historical cohort data that include detailed environmental parameters, comprehensive infection monitoring (especially for mild/asymptomatic cases), and precise clinical outcomes.

Confounding Factor Control: Many unmeasurable or uncontrollable confounding factors (e.g., socioeconomic status, differences in individual protective behaviors, healthcare resource availability) may exist between historical cohorts, all of which can influence the interpretation of results.

Limitations of Causal Inference: Retrospective studies are observational in nature; while they can reveal associations, they cannot directly prove causality. Observed associations may be driven by other unidentified factors.

Despite these challenges, if well-designed, data-rich historical cohorts can be identified for careful comparative analysis, the results can still provide valuable clues for understanding the relationship between total environmental viral aerosol exposure and the progression of respiratory infectious lung disease, thereby indirectly supporting certain aspects of this hypothesis.

6. Discussion

This paper proposes a hypothetical model concerning the dual role and dynamic shift of respiratory aerosols and inflammatory exudates in the dynamics and progression of SARS-CoV-2 lung infection, offering a novel perspective on how respiratory pathogens disseminate within the host lung. This section will delve into the importance of this hypothesis, reflect on current research paradigms, and discuss its potential applications in public health practice, as well as its existing limitations.

6.1 Reflection on Existing Research Paradigms and Lessons from Cholera Research

Looking back at the history of infectious disease research, John Snow's pioneering work on the London cholera epidemic in the 19th century offers a profound lesson for understanding current research on respiratory infectious diseases. At that time, the mainstream scientific community widely accepted the "miasma theory," believing that diseases were transmitted through "bad air." This, to some extent, parallels the current tendency in respiratory infectious disease research to overly focus on microscopic biological mechanisms (e.g., viral replication efficiency within cells, molecular pathways of host immune responses) while relatively neglecting the role of macroscopic physical environments and physiological media in the dynamic in-vivo dissemination of pathogens. Snow, through meticulous epidemiological investigation, ultimately traced the source of cholera transmission to contaminated water, completely overturning the "miasma theory" and leading to significant advancements in public health.

Similarly, this hypothesis attempts to expand our understanding of respiratory infectious lung infections from merely focusing on the "microscopic battlefield" within cells to viewing the lung as an open, dynamic physical space where aerosols and exudates serve as crucial physical media, dictating viral seeding and spread. For instance, some macroscopic intervention studies, such as "Nasal irrigation efficiently attenuates SARS-CoV-2 Omicron infection, transmission and lung injury in the Syrian hamster model [14]" and "Saltwater Gargling May Help Avoid COVID Hospitalization [15]," have preliminarily shown that simple physical methods like nasal irrigation and saltwater gargling can effectively reduce viral load and alleviate disease severity. These research findings align well with this hypothesis's emphasis on "oropharyngeal viral load peaks driving self-aerosol-mediated early intrapulmonary expansion," indicating the effectiveness of intervening in viral physical transmission by improving the local microenvironment. However, such macroscopic and easily scalable interventions have not yet received sufficient attention and in-depth mechanistic exploration in mainstream research commensurate with their potential impact.

6.2 Core Arguments, Relative Consensus, and Key Unverified Points of This Hypothesis

The core argument proposed in this paper is that aerosols play a critical role in initial seeding and early expansion, while after lung inflammation reaches a certain level and exudates form (i.e., reaching a dynamic equilibrium point), fluid transport then dominates the large-scale spread of the virus within the lungs.

Within this hypothetical framework, some arguments are based on existing knowledge and logical deduction, exhibiting high rationality, and can be considered relatively consensual or less controversial foundational parts of this hypothesis:

Aerosol-mediated initial seeding: Refers to the process by which external infectious sources transmit viruses to the host lung via aerosols, establishing initial infection foci. For example, in scenarios where individuals with active infection spend prolonged periods in the same confined space (e.g., family members sleeping together, dormitory residents), especially during the early stages of exposure (e.g., the first night or a longer time window), the inhalation of virus-laden aerosols undoubtedly plays a major role in initiating lung infection. This point, though still requiring more direct viral tracing studies, aligns with the basic principles of aerosol transmission and numerous epidemiological observations, thus having relatively less controversy at a logical level.

Inflammation-induced exudate-mediated widespread dissemination in later stages: When severe lung inflammation occurs, leading to increased capillary permeability and the formation of large amounts of virus-rich exudate, these fluids become effective vehicles for the physical dissemination of the virus between alveoli and airway branches, thereby driving rapid lesion fusion and diffuse progression. This mechanism also exhibits strong logical consistency and can explain the dynamic evolution characteristics of CT images in severe pneumonia, thus also falling into the less controversial category.

However, this hypothesis also contains several key aspects that require in-depth research and experimental validation, which constitute the main unverified points or potential controversial foci:

Definition, precise characterization, and critical "timing of arrival" of the "dynamic equilibrium point": In this hypothesis, the "dynamic equilibrium point" initially refers primarily to the critical point where the dominant mechanism of pathogen diffusion in the lung shifts from aerosols to inflammatory exudates. However, more broadly, it can also be understood as the transition point from the initial stage, primarily dominated by external aerosol-mediated seeding, to a stage where a combination of other diffusion mechanisms (including potential self-aerosol reseeding, cell-to-cell spread, and ultimately exudate transport) begins to dominate. For the overall explanation of pathogen diffusion, the latter understanding may have a broader consensus basis. Regardless of the chosen definition, the specific biological or physical markers of this "dynamic equilibrium point" (e.g., specific inflammatory cytokine level thresholds, initial exudate volume, or specific imaging transition characteristics), and the specific factors influencing its appearance (e.g., initial viral exposure dose, rate and intensity of host immune response, underlying health conditions, etc.) still need precise quantification. The core controversy lies in the certainty and individual variability of its "timing of arrival": For example, the aforementioned external aerosol initial seeding might dominate for "the first night or a longer time window," meaning this equilibrium point might appear approximately 1, 2 days, or even later after infection. This significant uncertainty in timing and the potential for large individual differences directly affect the judgment of the relative importance of subsequent different diffusion mechanisms (especially the role of self-aerosols), making it a key issue that urgently needs to be clarified through meticulous clinical dynamic observation and rigorous experimental research.

"Whether" and "under what conditions" self-aerosol reseeding plays a significant role in early expansion: This hypothesis postulates that, during a specific window period when oropharyngeal viral load peaks, but pulmonary specific immune barriers are not yet fully established, and large-scale inflammatory exudation has not yet occurred, virus-laden aerosols exhaled and re-inhaled by the host may drive the early multifocal dissemination of infection foci and the expansion of the infection area within the lungs. The core controversy here is "whether this mechanism can actually occur and produce a significant effect that surpasses other early mechanisms": If the "timing of arrival" of the aforementioned "dynamic equilibrium point" (especially the moment when other non-aerosol-dependent diffusion modes, such as local cell-to-cell spread or early micro-exudate, begin to play a significant role) occurs significantly earlier than the time when the host itself can produce peak viral aerosols capable of sufficient intrapulmonary dissemination, then the role of "self-aerosol reseeding" in early expansion may be very limited or even negligible. Therefore, determining if and when this mechanism can play a role, and its actual contribution (if present and significant) to the total viral load growth and lesion expansion in the lung, is one of the core innovative inferences of this hypothesis and the most critical question that needs careful validation through ingenious experimental designs (such as the ideal experiment concept described in Section 4.2.2) and quantitative comparative studies with other early diffusion mechanisms (such as local cell-to-cell spread).

Although some aspects of the hypothesis (especially regarding the critical driving role of self-aerosols in early expansion and their quantitative contribution) require further rigorous validation, its less controversial parts, particularly the assertion about the importance of external aerosol initial seeding in the establishment phase of infection, already possess theoretical significance for guiding current epidemic prevention practices. This suggests that existing respiratory infectious disease prevention strategies could be appropriately adjusted and strengthened based on this. For example, during high seasons for respiratory infectious diseases, for all indoor places with high population density (even just two people in a bedroom or collective dormitory), relatively confined spaces, poor ventilation, and where people need to stay for extended periods, effective measures should be particularly emphasized and adopted (such as improving ventilation, continuously operating high-efficiency air purification equipment, etc.) to, where feasible, minimize the concentration of pathogenic aerosols in the environment, thereby preventing or significantly reducing the occurrence of initial infection events and the initial inhaled viral dose. This may not only reduce the probability of individual infection but also potentially alleviate the severity of subsequent disease development by reducing the initial "seeding" amount.

6.3 Practical Implications: Individual Action and Bidirectional Protection – A New Perspective for Optimizing Respiratory Health Strategies

Based on this hypothesis, the intervention measures we advocate can be seen as a natural extension and refinement of existing respiratory infectious disease prevention strategies. Current prevention strategies primarily focus on preventing or controlling pathogen transmission between people and their spread in traditional public places. On this basis, this paper further emphasizes the often-overlooked aspect of "individual-to-self transmission." This means that even in seemingly private spaces, such as an individual's bedroom during sleep, consideration should be given to reducing the re-inhalation and local accumulation of self-exhaled aerosols through moderate ventilation (e.g., keeping the bedroom door open during sleep to promote air exchange, or using a low-power HEPA air purifier if necessary). Similarly, dormitories and homes, though not traditionally considered large "public" places, are microenvironments that require particular attention due to members sharing limited space for extended periods, to reduce internal transmission and the risk of individual self-re-inhalation.

From the perspective of individual willingness to prevent infection, existing prevention concepts mostly emphasize preventing external pathogens from entering oneself. This hypothesis further points out that while actively taking measures to prevent external pathogen entry, if one can consciously control the accumulation and re-inhalation of self-generated pathogens in the local microenvironment (i.e., controlling "individual-to-self transmission"), then this behavior also significantly reduces the individual's ability to transmit pathogens to the external environment. This creates a "for all, by all" (literally "everyone for me (by reducing self-re-inhalation to protect personal health), I for everyone (by reducing environmental release to protect others)") win-win situation. From this perspective, even if the core mechanism of this hypothesis regarding self-aerosol re-inhalation accelerating disease worsening has not been fully confirmed, the concept of improving individual microenvironments and strengthening respiratory hygiene advocated in this paper itself still holds significant positive implications for enhancing overall epidemic prevention effectiveness, as it undoubtedly reduces the probability of pathogen transmission between people.

Driven by this bidirectional benefit awareness, more people who encounter the content of this paper can become important participants in promoting the validation and application of this theory through simple understanding and active daily practices. By raising public awareness of aerosol transmission and the potential risks of self-re-inhalation, and by encouraging more proactive ventilation and air purification measures in homes, schools, offices, and other environments (especially when individuals with respiratory symptoms are present), we will be able to generate a large amount of observational data and experience. These real-world, widely participated practices and observations, while not constituting rigorous scientific experiments, can accumulate evidence that strongly supports or challenges the practical relevance of this hypothesis and accelerates the acceptance and adoption of this new paradigm by the scientific community and public health agencies.

However, if this hypothesis can be further confirmed and applied to practical guidance, it must also be emphasized that a balanced and cautious approach is necessary. For example, the seemingly simple advice of "opening windows for ventilation in winter" becomes more complex to apply if the hypothesis is supported, requiring careful consideration. Opening windows for extended periods on cold winter nights to achieve maximum ventilation, while theoretically helpful in reducing aerosol accumulation, may also lead to excessively low indoor temperatures, increasing discomfort and even inducing other health problems like colds, where the potential health risks might outweigh the benefits of epidemic prevention. This suggests that when promoting any intervention measures based on this hypothesis, it is crucial to fully consider multiple factors such as season, specific environmental conditions, individual physiological status, cultural habits, and economic feasibility, striving to achieve a scientific balance between improving air quality and maintaining a comfortable and healthy living environment, avoiding short-sightedness.

6.4 Limitations and Future Research Directions of the Hypothesis

The hypothetical model proposed in this paper regarding the dynamic diffusion of respiratory infectious diseases within the host lung provides a novel framework focused on physical transmission media and their dynamic shifts for understanding the pathophysiological process. Although we have thoroughly analyzed the core arguments, relatively consensual foundations, and main points of contention requiring validation in Section 5.2, as a theoretical model still under development, it currently exhibits several important limitations. These limitations directly point towards key breakthrough directions for future research.

6.4.1 Deep Integration of Complex Host Immune Responses and Pathophysiological Processes:

To highlight the physical transport role of aerosols and exudates, this model currently significantly simplifies the complex dynamics of the host immune system and its influence on pathogen amplification. This simplification poses additional challenges, particularly when discussing the timing of peak lung pathogen load. For patients with mild disease, lung infection is generally less severe, making precise measurement of peak viral load almost impossible; furthermore, the host immune system may clear pathogens earlier and more effectively, which further increases the difficulty of determining peak lung viral load without adequately considering the role of the immune system. In light of this, this paper primarily bases its discussion of peak lung pathogen load on cases involving severe patients, focusing on inferring viral load dynamics using indirect evidence such as CT imaging changes, to compensate for the limitations of direct viral load measurement in mild cases.

However, the host's immune response (including the rapid initiation of innate immunity, the establishment and effector mechanisms of adaptive immunity, the fine regulation of cytokine networks, and potential immunopathological damage and tissue repair processes) is deeply intertwined with and causally linked to viral replication, clearance, and physical diffusion within the lungs. Future research should aim to construct multi-scale, multi-compartment computational models capable of integrating these critical immune parameters, and calibrate and validate them using clinical cohort data and animal experimental data, to more comprehensively understand how the immune system dynamically regulates the efficiency of physical diffusion, defines the formation and properties of exudates, and ultimately determines the clinical phenotype and outcome of the disease.

6.4.2 Comprehensive Consideration of Other Concurrent Diffusion Pathways, Host Heterogeneity, and Environmental Factors:

Although this hypothesis focuses on aerosols and exudates as two key physical media, direct viral transmission between infected cells, potential minor dissemination through the intrapulmonary lymphatic network or circulatory system, and the efficiency of local non-directional diffusion of the virus within the alveolar microenvironment, all remain important components contributing to the overall complex picture of lung infection. Their relative contributions may vary at different stages of the disease or among different individuals. Furthermore, how intrinsic host heterogeneity (such as genetic background differences, subtle anatomical and physiological variations in the lung, pre-existing respiratory diseases, and even daily breathing patterns and depths) and external microenvironmental factors (such as air humidity, temperature, and pollutant levels affecting aerosol behavior and host respiratory tract conditions) finely regulate aerosol deposition patterns, the functional status of the mucociliary clearance system, and the intensity and type of inflammatory response, thereby influencing the various dynamic stages described in this hypothesis, also require more systematic and in-depth research.

6.4.3 Specific Pathogen Applicability:

This paper primarily focuses on SARS-CoV-2, but it should be noted that different respiratory infectious disease pathogens exhibit significant differences in their biological characteristics (e.g., virus size, stability, host cell tropism, replication kinetics, type and intensity of inflammatory response induced), which may lead to different physical transmission and diffusion patterns within the host lung. Therefore, the specific dynamic mechanisms and equilibrium points in this hypothesis may differ for other respiratory pathogens (such as influenza virus, respiratory syncytial virus, etc.) and require targeted research and validation.

To overcome these limitations and promote the continuous refinement and validation of the hypothesis, future research strategies should emphasize deeper interdisciplinary integration and collaborative innovation, effectively combining epidemiological surveys, clinical translational medicine research, basic virology and immunology experiments, biofluid mechanics and physical modeling, and advanced computational biology and artificial intelligence analysis. Particularly, new technologies and methods aimed at directly observing or indirectly inferring the dynamic behavior of viral particles (or their highly credible surrogates) in a near-real pulmonary microenvironment (with full consideration and strict adherence to medical ethics) should be prioritized for development and investment. This will be crucial for directly validating or modifying the core mechanistic links of this hypothesis.

7. Conclusion

This paper proposes a hypothetical model for the dynamic diffusion of respiratory infectious diseases (taking SARS-CoV-2 as an example) within the host lung. The core idea is that aerosols play a crucial role in initial seeding and early expansion, while inflammatory exudates formed after inflammation dominate the transport of viruses in large-scale lesion spread. This dynamic equilibrium model offers a new physical and dynamic analytical perspective for understanding the complex process of COVID-19 pneumonia's rapid evolution from focal to diffuse.

Although real-time, high-resolution direct validation of the microscopic dynamic processes described in this hypothesis in living humans faces significant ethical and technical challenges, existing clinical observational data, viral kinetic studies, and epidemiological pattern analyses all provide consistent indirect evidence for this hypothetical model. Our conceptualized ideal experimental designs, as well as retrospective associative analysis methods based on historical cohort data, while presenting implementation difficulties and inherent limitations, point the way for future research to further examine the impact of viral aerosol exposure on disease progression through environmental intervention measures.